\documentclass[sigconf]{acmart} 
\usepackage{graphicx}
\usepackage{subcaption}
\usepackage{hyperref}
\usepackage{url}
\usepackage{algorithm}
\usepackage{algorithmic}

\usepackage{amsmath, amsfonts}
\usepackage{bbold}

\usepackage[textsize=tiny,textwidth=1.4cm]{todonotes}

\AtBeginDocument{%
  }

\copyrightyear{2026}
\acmYear{2026}
\setcopyright{cc}
\setcctype{by}
\acmConference[ICAIF '26]{7th ACM International Conference on AI in Finance}{November 14--17, 2026}{Milan, AA, Italy}
\acmBooktitle{7th ACM International Conference on AI in Finance (ICAIF '26), November 14--17, 2026, Milan, AA, Italy}
\acmDOI{10.1145/3858063.3858680}
\acmISBN{979-8-4007-3087-0/2026/11}

\begin{document}

\title{Linguistic Passiveness in Earnings Calls and Stock Returns}

\author{Lynn Tao}
\email{lynnyl.tao@mail.utoronto.ca}
\orcid{0009-0000-3856-2146}
\affiliation{%
  \institution{University of Toronto}
  \city{Toronto}
  \state{Ontario}
  \country{Canada}
}
\author{Yuntao Wu}
\email{winstonyt.wu@mail.utoronto.ca}
\orcid{0009-0006-6269-1603}
\affiliation{%
  \institution{University of Toronto}
  \city{Toronto}
  \state{Ontario}
  \country{Canada}
}
\author{Charles Martineau}
\email{charles.martineau@rotman.utoronto.ca}
\orcid{0000-0002-6896-184X}
\affiliation{%
  \institution{University of Toronto}
  \city{Toronto}
  \state{Ontario}
  \country{Canada}
}
\author{Vincent Grégoire}
\email{vincent.gregoire@hec.ca}
\orcid{0000-0002-1230-5026}
\affiliation{%
  \institution{HEC Montréal}
  \city{Montréal}
  \state{Québec}
  \country{Canada}
}
\author{Andreas Veneris}
\email{veneris@eecg.toronto.edu}
\orcid{0000-0002-6309-8821}
\affiliation{%
  \institution{University of Toronto}
  \city{Toronto}
  \state{Ontario}
  \country{Canada}
}

\renewcommand{\shortauthors}{L. Tao, Y. Wu, C. Martineau, V. Grégoire, A. Veneris}

\begin{abstract}
We study linguistic passiveness as the fraction of passive-voice sentences in earnings conference call Q\&A sessions and show it is informative for equity markets. Using 105,642 U.S. calls from 2008 to 2024, passiveness is negatively associated with announcement-day returns among positive-surprise firms: returns fall from 4.38\% to 2.74\% across passiveness quintiles. This announcement-day discount fully reverses over the subsequent quarter, indicating a temporary mispricing when news is good. For negative-surprise firms, passiveness has no announcement-day effect but predicts negative long-run returns, consistent with investor under-attention when salient bad news dominates. Passiveness remains informative after controlling for sentiment and standard readability proxies, suggesting that linguistic passiveness captures a distinct yet important dimension of managerial communication.
\end{abstract}




\maketitle

\section{Introduction}

Recent advances in large language models and textual analysis give researchers the ability to extract information from financial disclosures. Yet much of this progress relies on models whose predictions are difficult to interpret: it remains unclear which features of managerial language are most informative. An alternative is to study specific, linguistically grounded dimensions of communication whose meaning is precisely understood. 
In this paper, we examine one such dimension: whether managers use passive or active language during earnings conference calls, and whether that choice contains information relevant for financial markets.

Language conveys information not only through \emph{what} is said, but also through \emph{how} it is expressed. Prior research on managerial communications has focused primarily on content -- what managers discuss, and sentiment -- whether they describe those topics positively or negatively \cite{Loughran-McDonald-11,garcia2023colour,ecc-analyzer,earnings1}. 
We instead study syntactic framing: how managers structure statements when describing corporate outcomes. Consider the sentences ``We revised the forecast'' and ``The forecast was revised.'' Although both statements convey the same information and sentiment, the latter (passive language) places more emphasis on the outcome and less emphasis on who is responsible for the action.
Research in linguistics and psychology shows that such linguistic choices influence perceptions of responsibility and accountability \cite{passive-1999,simchon2023agency,fannes2022crisis}.
Because earnings conference calls involve managerial communication under uncertainty, accountability, and performance evaluation, variation in the use of passive language may signal how managers frame their role in the firm performance.

We measure passiveness as the fraction of passive sentences in the Q\&A portion of each call, identified using PassivePy \cite{passivepy}, a validated rule-based tool that detects passive constructions with 98\% agreement with human coders (with 2.7\% false-positive rate and 2.2\% false-negative rate). We apply this measure to 105,642 U.S. conference call events from 2008 to 2024, matched to earnings and return data from LSEG IBES and CRSP.

This paper documents five main findings. First, passiveness rises sharply during recessions and periods of market stress and declines gradually as equity markets grow. 
Sectors such as Utilities, Health Care, and Energy exhibit the highest average passiveness, consistent with the heightened regulatory scrutiny and litigation exposure in these industries: when drivers of firm performance are determined by external parties---regulators, commodity markets, approval bodies---passive constructions are more likely. Second, passiveness negatively predicts announcement-day abnormal returns, with the effect concentrated among firms reporting positive earnings surprises. Among the highest-surprise firms, average announcement-day returns fall from 4.38\% to 2.74\% moving from the lowest to the highest passiveness quintile, suggesting that investors discount favorable earnings news when managers simultaneously use language that distances them from the reported results. This announcement-day discount, however, fully reverses over the subsequent quarter, suggesting that the initial price reaction to passiveness is temporary when news is good. Third, for negative-surprise firms, passiveness has no announcement-day effect but predicts negative returns over the subsequent 61 trading days; consistent with investor under-attention to subtle linguistic cues when earnings news is unfavorable.  Fourth, these return effects survive controls for the \citet{garcia2023colour} sentiment measure and standard readability proxies, and are not explained by a fixed managerial communication style. 
Finally, recent market declines and firm-level underperformance are the strongest predictors of elevated passiveness.

Our paper contributes to the financial text-analysis literature by introducing grammatical passiveness as a transparent and economically interpretable measure of managerial communication. Unlike features extracted from increasingly complex language models, passiveness has a clear linguistic interpretation rooted in how speakers frame outcomes and responsibility. This transparency allows us to identify a specific aspect of managerial language that predicts investor reactions and future returns.



\section{Related Work}

A large body of research applies natural language processing to financial texts to extract information beyond traditional accounting measures. Early work relies on word-count and dictionary-based approaches to quantify the tone of corporate disclosures.
\citet{Loughran-McDonald-11} develop a financial domain-specific sentiment lexicon from 10-K filings and show that negative word counts predict returns, volatility, and fraud. \citet{garcia2023colour} extend this approach using a multinomial inverse regression framework to construct a return-related sentiment dictionary that yields stronger predictive power for stock returns.
More recent work leverages neural language models to capture richer semantic features. \citet{bert} introduce BERT, a pre-trained bidirectional transformer architecture that has become the backbone of numerous downstream financial NLP tasks, including the domain-adapted FinBERT \cite{finbert}. \citet{ecc-analyzer} combine large language models with retrieval-augmented generation to extract predictive signals from earnings call transcripts for volatility forecasting. Bridging dictionary- and neural-network-based methods, \citet{earnings1} show that the content of earnings press releases extracted via both bag-of-words and BERT-based embeddings is as informative as the earnings surprise for predicting announcement-day returns. Common to this literature is a focus on what managers communicate: the topics discussed and the positive or negative tone attached to them.

A distinct stream of research examines the framing of managerial communication. A well-established finding from social psychology is that individuals tend to attribute successes to their own actions while attributing failures to external circumstances, i.e., a pattern known as self-serving attribution bias \cite{bradley1978self,zuckerman1979attribution}. This pattern has been documented in corporate settings as well. \citet{doukas2007acquisitions} find evidence of self-attribution bias among acquiring managers. More recently, \citet{earnings1} apply topic modeling to earnings press releases and document that managers emphasize internal drivers when reporting positive outcomes and shift blame externally when results disappoint. \citet{huang2014tone} show that managers strategically manage the tone of earnings disclosures to influence investor perceptions, a practice they term ``tone management.'' While this literature shows that managerial communication is shaped by strategic incentives, it has paid less attention to the grammatical structure through which managers communicate these messages.

One aspect of grammatical structure that has received considerable attention in linguistics and social psychology is passive voice. Passive voices have long been recognized as a mechanism for de-emphasizing agency and shifting attention from actors to outcomes. \citet{passive-1999} show experimentally that passive voice reduces attributions of responsibility for described behaviors, independent of group membership. \citet{simchon2023agency} provide large-scale computational evidence that passive language suppresses the linguistic expression of personal agency, while \citet{fannes2022crisis} demonstrate that passive voices shape how responsibility is attributed in crisis communication. Collectively, this literature establishes that grammatical voice carries meaning beyond lexical content. To the best of our knowledge, passive voice has not been systematically examined as a predictor of stock return reactions to financial disclosures.

\section{Data and Methodology}

\subsection{Conference Call and Financial Data}

\begin{figure*}[!htb]
  \centering
  \begin{subfigure}[b]{0.3\linewidth}
  \centering
  \includegraphics[width=\linewidth]{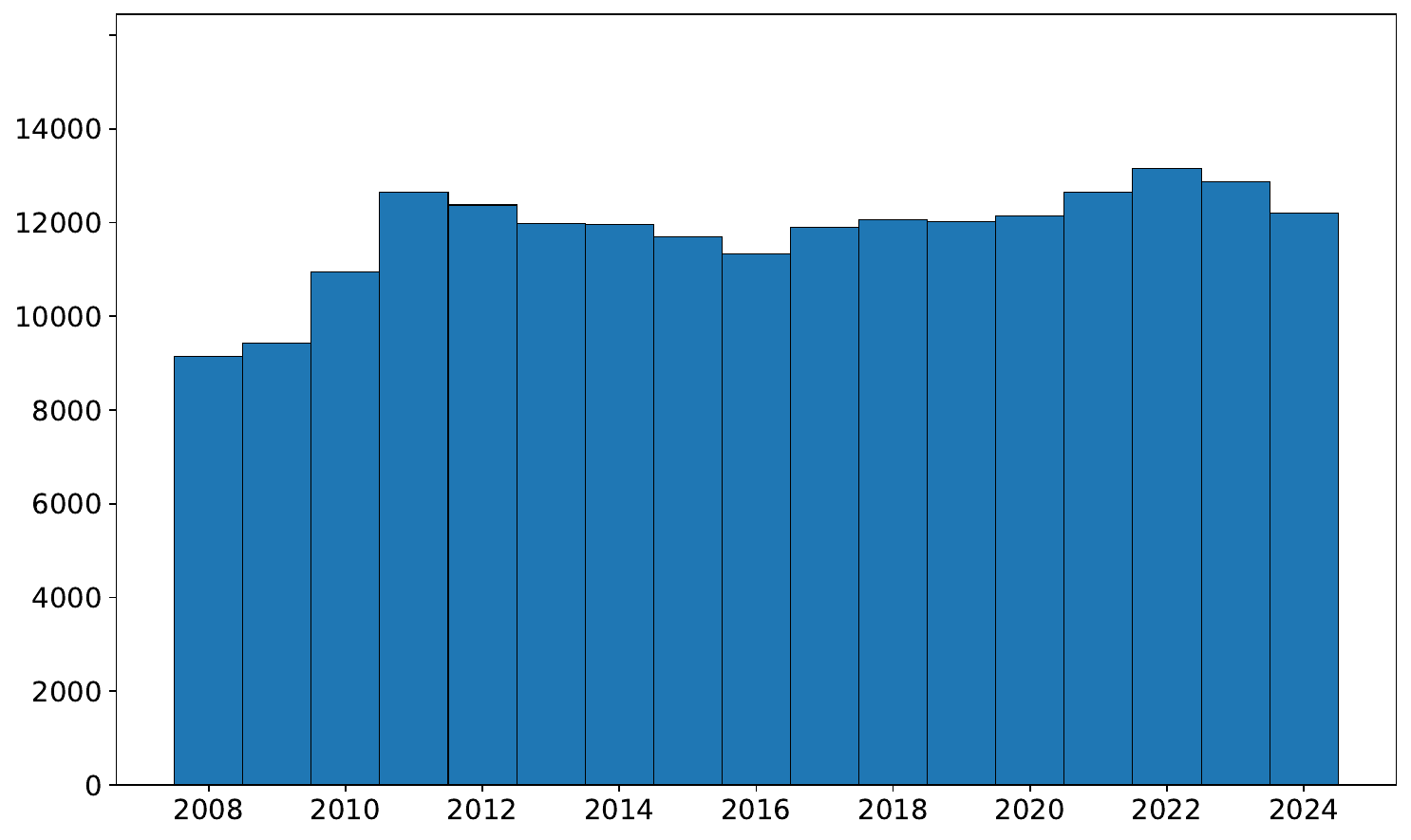}
  \caption{Number of Calls \label{fig:stats-calls}}
  \end{subfigure}
  \hfill
  \begin{subfigure}[b]{0.3\linewidth}
  \centering
  \includegraphics[width=\linewidth]{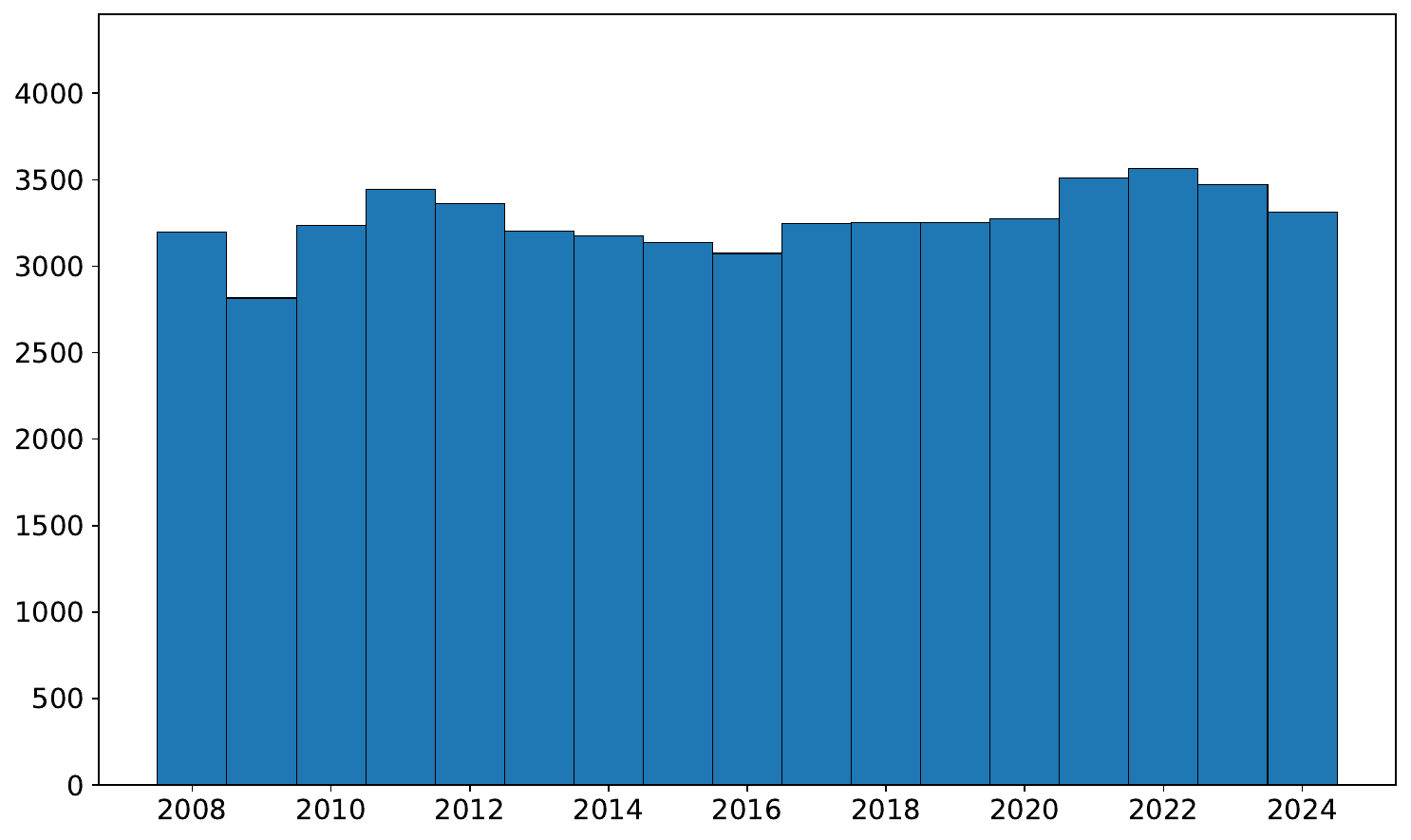}
  \caption{Number of Stocks \label{fig:stats-stocks}}
  \end{subfigure}
  \hfill
  \begin{subfigure}[b]{0.3\linewidth}
  \centering
  \includegraphics[width=\linewidth]{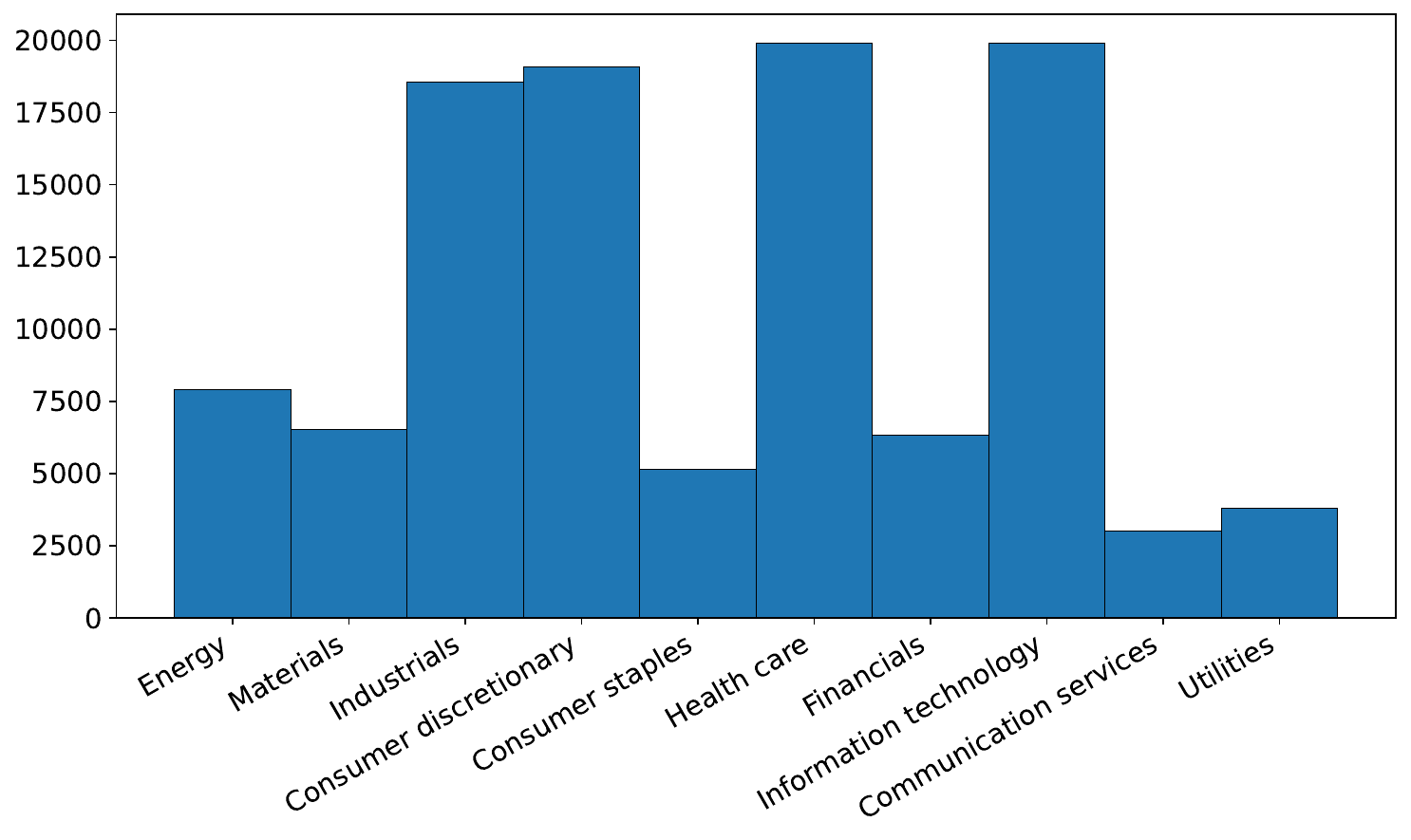}
  \caption{Number of Calls per Sector \label{fig:stats-calls-sectors}}
  \end{subfigure}
  \caption{General Statistics \label{fig:statistics}}
  \Description{General Statistics}
\end{figure*}

We retrieve U.S. corporate conference call transcripts from S\&P Global, and focus on the Q\&A portion of each call, as analyst questions elicit largely unscripted responses that provide a more natural setting for observing spontaneous managerial language use than the scripted prepared remarks. Conference calls occurring after market close (4:00 p.m. \ ET) are assigned an event date on the next trading day, whereas calls occurring during market hours are assigned to the same trading day. Accordingly, a conference call occurring between 4:00 p.m. on trading day $t$ and 4:00 p.m. on trading day $t+1$ is associated with the announcement-day return from day $t$ to day $t+1$. In contrast, a conference call occurring before 4:00 p.m. on trading day $t$ is assigned to day $t$ and is therefore associated with the return from day $t-1$ to day $t$. 
The raw sample contains 200,532 Q\&A sessions from January 2008 through December 2024, covering an average of 3,266 firms per year. Figures~\ref{fig:stats-calls} and \ref{fig:stats-stocks} present the number of conference calls and firms by year. We obtain earnings announcement dates and times, analyst forecasts, and reported earnings per share (EPS) from LSEG IBES Academic. Sector classifications are identified using GICS codes in Compustat, while daily stock returns are obtained from the CRSP Daily Stock Files. We merge the data from S\&P Global transcripts with IBES, CRSP, and Compustat using the GVKEY and PERMNO linking tables provided by WRDS. 

Following \cite{martineau2021rest}, we define the earnings surprise for firm $c$ at announcement date $\tau$ as
\begin{equation}
	\text{Surprise}_{c,\tau} =  \frac{\text{EPS}_{c,\tau}-E_{\tau-1}[\text{EPS}_{c,\tau}]}{P_{c,\tau-5}},
\end{equation}
where $\text{EPS}_{c,\tau}$ denotes the reported EPS for the announcement, and $E_{\tau-1}[\text{EPS}_{c,\tau}]$ is the expected EPS proxied by the consensus analyst forecast, defined as the median of each analyst’s most recent EPS forecast issued in the 90 days preceding the announcement date.
We scale the surprise by the firm’s stock price $P_{c,\tau-5}$ five trading days before the announcement, and winsorize at the 1st and 99th percentiles per year.

For each firm $c$, we compute announcement-day abnormal returns and long-horizon buy-and-hold abnormal returns (BHARs) over trading day windows $[t_1,t_2]$:
\begin{align}
  \text{AR}_{c, \tau} &= \text{Ret}_{c,\tau} - \text{Ret}_{p,\tau}\\
  \text{BHAR}[t_1, t_2] &= \prod_{t=t_1}^{t_2} (1+\text{Ret}_{c,t}) - \prod_{t=t_1}^{t_2} (1+\text{Ret}_{p,t}),
\end{align}
where $\text{Ret}_{c,t}$ is the close-to-close return, and the benchmark return $\text{Ret}_{p,t}$ is the return on a size- and book-to-market-matched Fama--French benchmark portfolio.

After merging conference call data with return and earnings information, we retain observations with valid earnings surprise and announcement-day abnormal returns. The reduction from the raw sample primarily reflects calls unmatched to IBES earnings announcements, observations with missing CRSP return data, and observations with missing analyst forecast data. The final sample contains 105,642 conference call events from 4,945 firms, corresponding to an annual average of 6,214 events from 2,147 firms.
Figure~\ref{fig:stats-calls-sectors} shows the distribution of the conference calls across the ten GICS sectors; the Real Estate sector is absent from the final sample because no conference calls from that sector remain after the data merge. 

\subsection{Passiveness of Text}

\begin{figure}[!htb]
  \centering
  \includegraphics[width=1\linewidth]{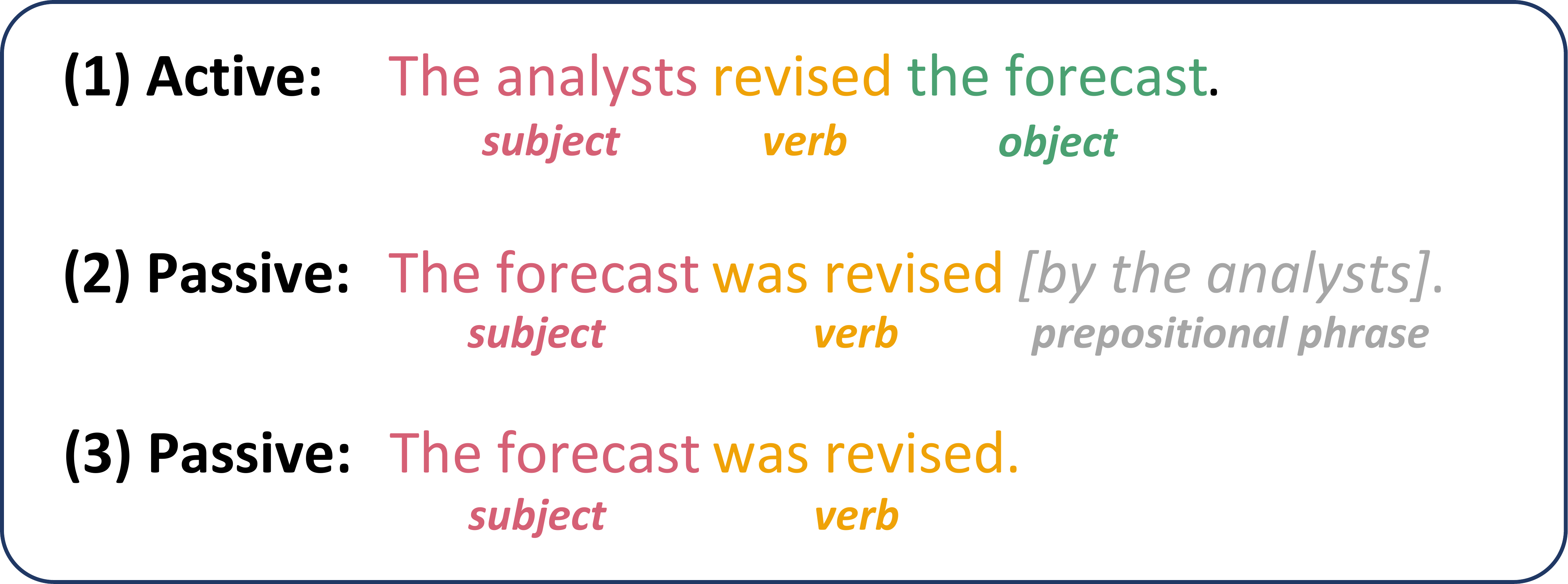}
  \caption{Illustration of active and passive sentence constructions}\label{fig:passive-ex1}
  \Description{Example}
\end{figure}

In everyday communication, speakers may choose between alternative grammatical constructions that describe the same event while emphasizing different aspects of it. One such choice is between active and passive sentence constructions. 
Figure~\ref{fig:passive-ex1} illustrates the difference between active and passive sentence. In the active sentence in (1), ``The analysts revised the forecast'', the sentence emphasizes the actor (``the analysts'') as the subject performing the action (``revis[ing] the forecast''). In contrast, the passive sentence in (2), ``The forecast was revised [by the analysts],'' shifts the attention towards the outcome (``the forecast'') rather than the actor. The phrase containing the actor, ``by the analysts'', may even be omitted as shown in (3), leaving the responsible party unspecified. Although both the active and passive constructions describe the same event, they differ in how responsibility and emphasis are communicated.

\begin{figure}[!htb]
  \centering

  \begin{subfigure}{0.48\linewidth}
    \centering
    \includegraphics[width=\linewidth]{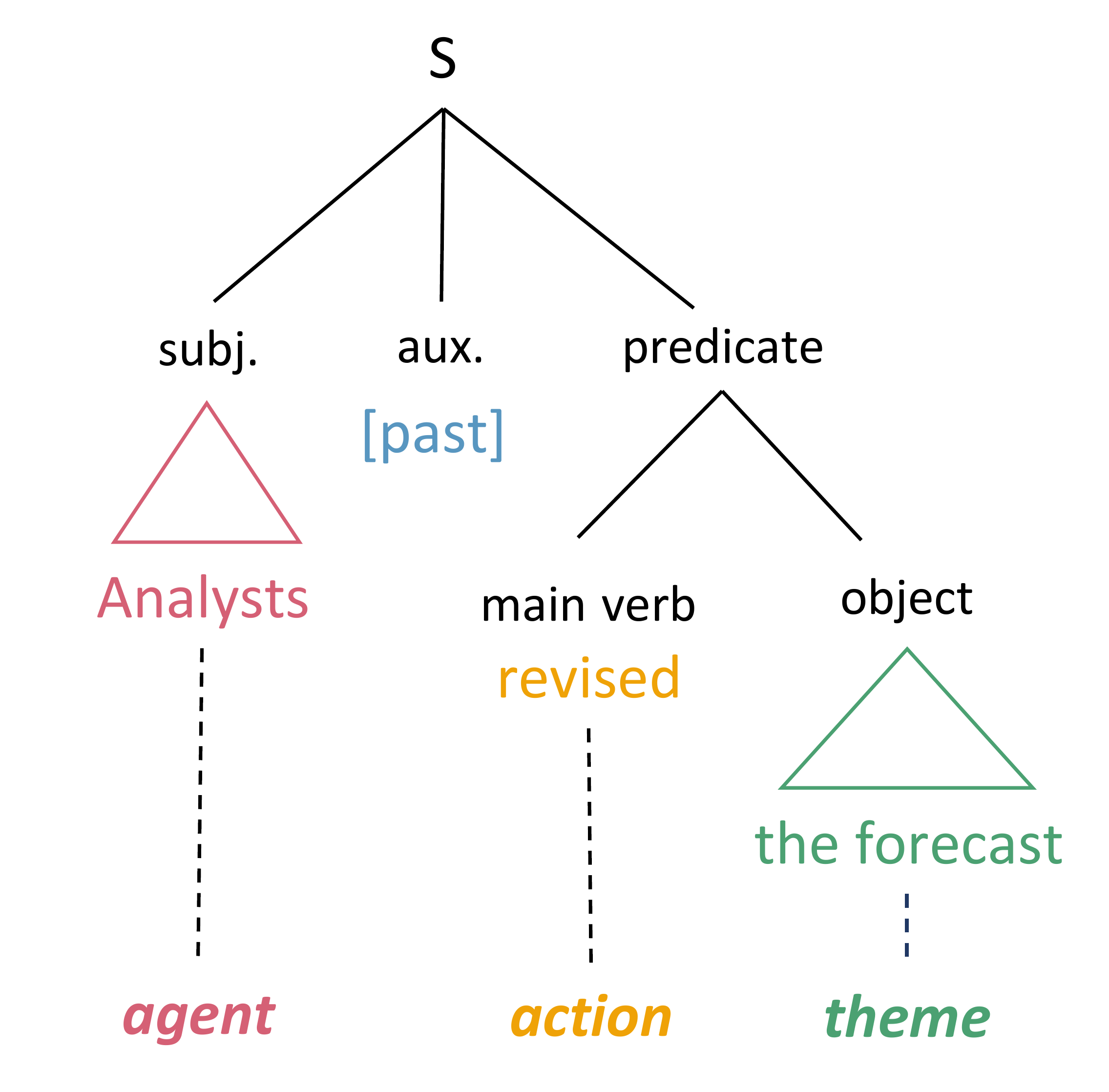}
    \caption{Active}
    \label{fig:tree-1}
    \Description{Example}
  \end{subfigure}
  \hfill
  \begin{subfigure}{0.48\linewidth}
    \centering
    \includegraphics[width=\linewidth]{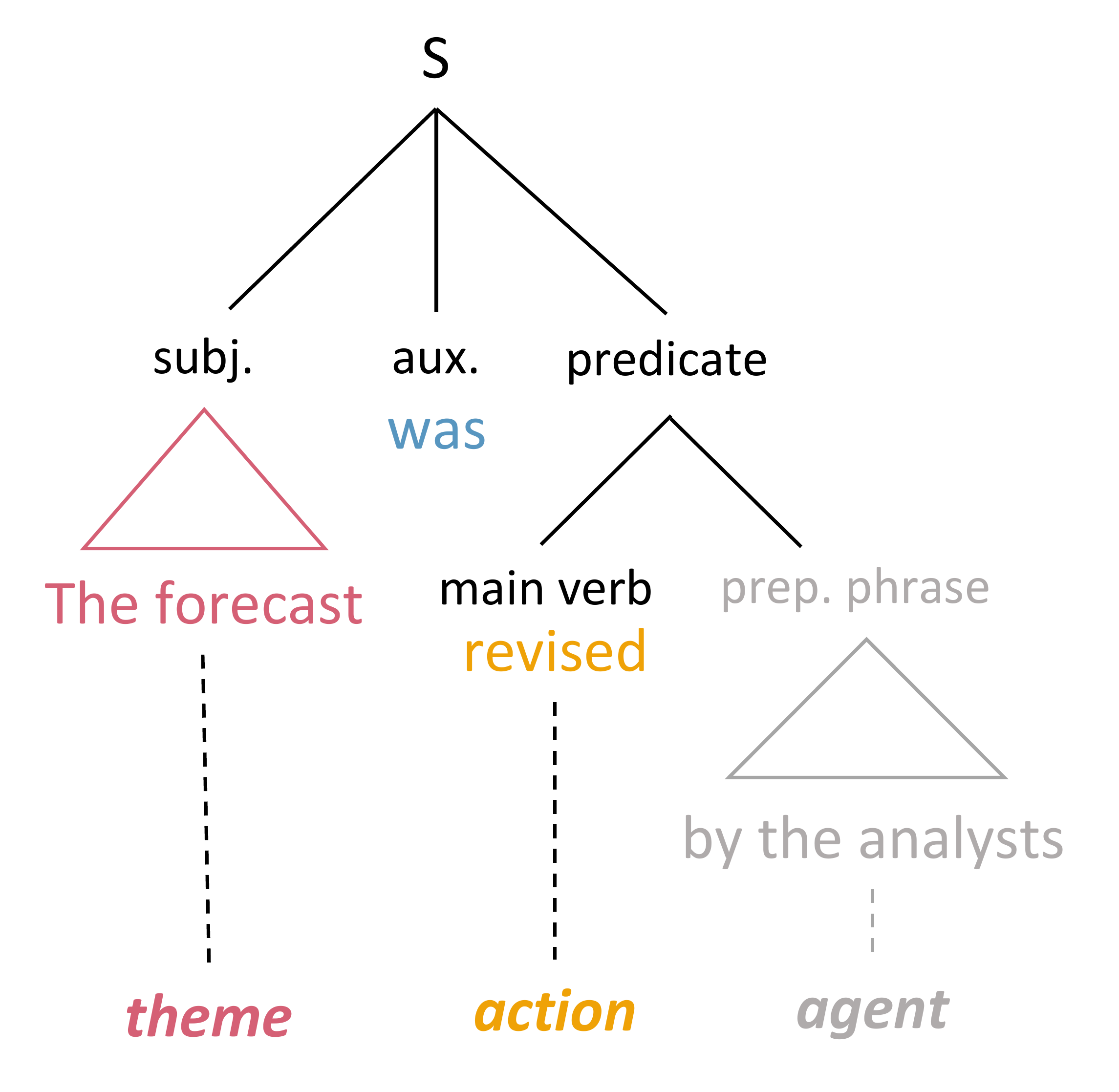}
    \caption{Passive}
    \label{fig:tree-2}
    \Description{Example}
  \end{subfigure}

  \caption{Syntactic trees of active and passive constructions}
  \label{fig:trees}
\end{figure}

More formally, passive voice is a grammatical construction in which the subject \textit{receives} an action rather than \textit{performs} it. In active constructions, the subject typically corresponds to the entity initiating the action (called the \textit{agent}), while the object corresponds to the entity affected by the action (called the \textit{theme}). Passive constructions reorganize these roles syntactically: the theme becomes the subject, while the original agent is demoted to an optional prepositional phrase or omitted entirely. Consequently, passive voice often backgrounds agency and foregrounds events or outcomes. The resulting change in grammatical relations is reflected in the syntactic trees shown in Figure~\ref{fig:trees}, where the subject corresponds to the theme in the passive rather than the agent in the active.

\begin{figure}[!htb]
  \centering
  \includegraphics[width=1\linewidth]{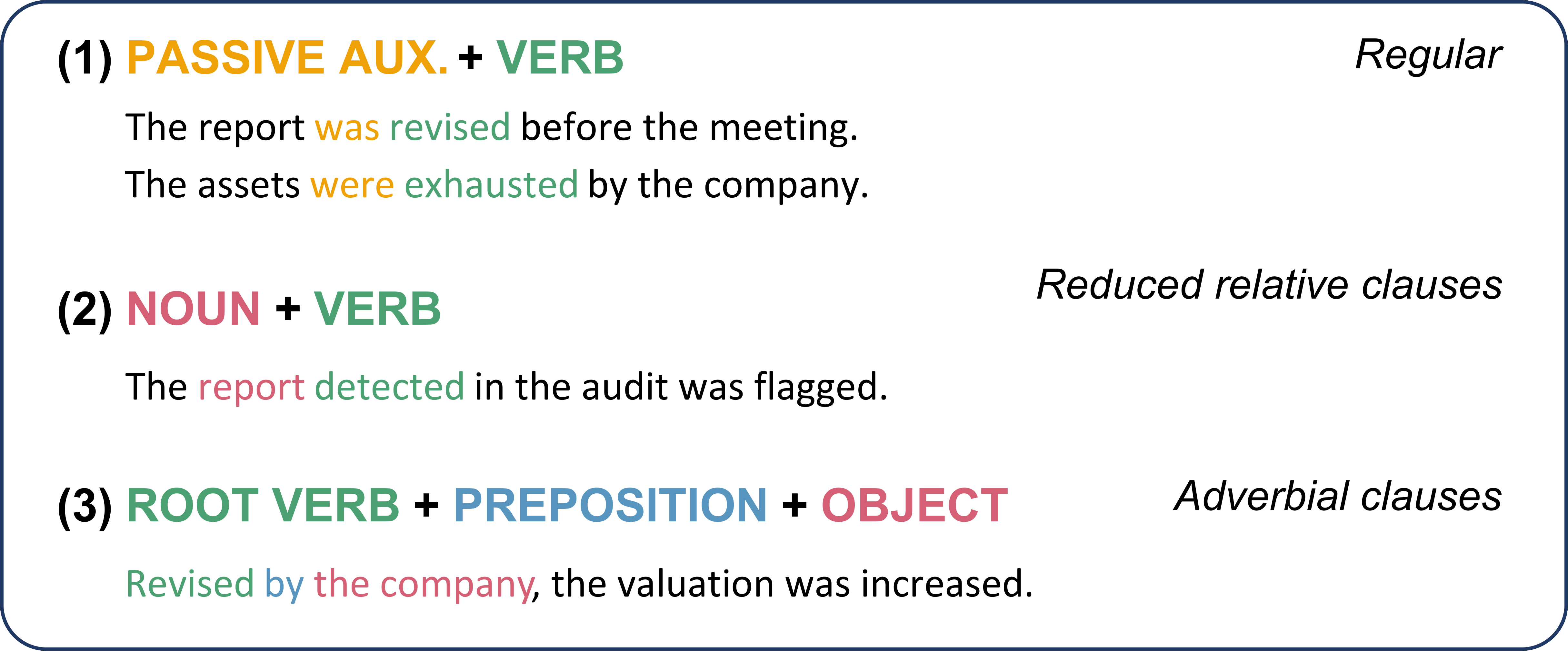}
  \caption{Simplified matching rules}\label{fig:ex-rules}
  \Description{Example}
\end{figure}

The PassivePy library \cite{passivepy} identifies passive constructions using a rule-based linguistic parsing approach built on the spaCy natural language processing framework \cite{spacy}. The library first applies Part-of-Speech (POS) tagging to assign grammatical categories to each token (e.g. nouns, verbs, adjectives, adverbs). Then, it performs dependency parsing to identify relationships between words in the context of the sentence. Using these linguistic features, PassivePy applies five pattern-matching rules to detect different types of passive constructions. These include, among others, auxiliary-based passives with a form of \textit{be/have} followed by a past participle verb, reduced relative clauses functioning as noun phrases, and adverbial clauses. The output is a binary classification indicating whether passive voice is present in a sentence. A simplified summary of the rules is shown in Figure~\ref{fig:ex-rules}. 

In this paper, we define ``passiveness" as the fraction of sentences containing passive structures for each conference call transcript:
\begin{equation}
  \text{Passiveness}=\frac{\# \text{ passive sentences}}{\# \text{ total sentences}}
\end{equation}
This measure captures the extent to which speakers rely on passive grammatical structures during conference call discussions.

\subsection{Sentiment Score}

To examine whether linguistic content and tone are related to passiveness, we construct sentiment scores using the dictionary developed by \citet{garcia2023colour}. This dictionary extends and refines the \citet{Loughran-McDonald-11} financial sentiment lexicon. The tokens in the dictionary are identified through a multinomial inverse regression framework \cite{mnir}. Words are classified as positive or negative based on whether their estimated loadings exhibit consistent association with stock returns across samples.

We first apply standard text cleaning to the transcripts that removes numerical values, URLs, and stopwords. We then vectorize using the unigram version of the dictionary, as bigram occurrences are relatively sparse in conference call transcripts and results of \citet{garcia2023colour} indicate that unigram has the strongest predictive power for returns. 
For each firm $c$ and announcement day $\tau$, we assign a sentiment score by polarity:
\begin{equation}
  \text{Sentiment}_{c,\tau} = \frac{N_{c,\tau}^+-N_{c,\tau}^-}{N_{c,\tau}^++N_{c,\tau}^-},
\end{equation}
where $N_{c,\tau}^{+}$ and $N_{c,\tau}^{-}$ denote the counts of positive and negative unigram occurrences in the transcript, respectively. 
Higher values indicate more positive language, while lower values reflect relatively negative sentiment.

\section{Empirical Results}

\subsection{Descriptive Evidence}
\begin{figure*}[!htb]
  \centering
  \begin{subfigure}[b]{0.49\linewidth}
  \centering
  \includegraphics[width=\linewidth]{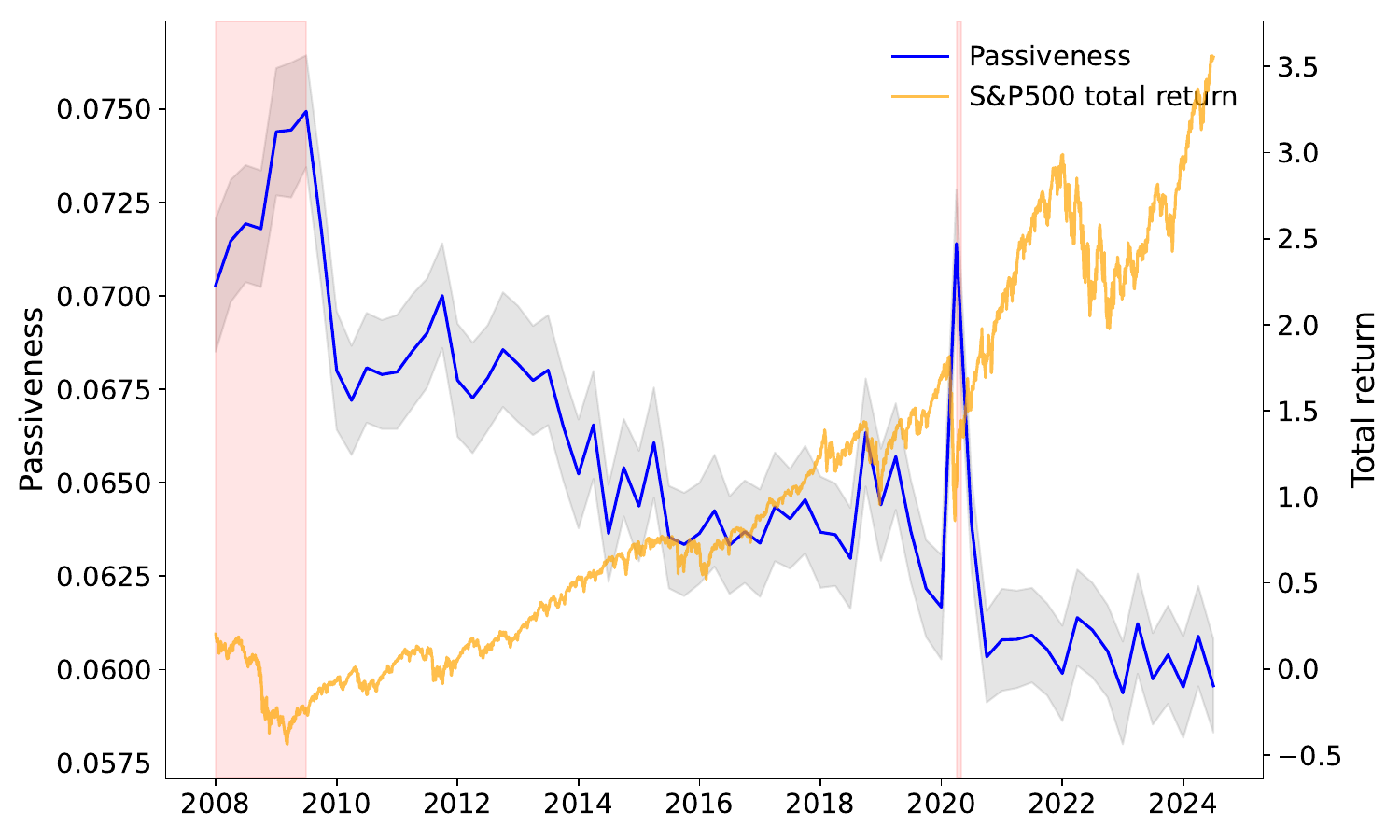}
  \caption{Average Passiveness \label{fig:timeseries-avg-passiveness}}
  \end{subfigure}
  \hfill
  \begin{subfigure}[b]{0.49\linewidth}
  \centering
  \includegraphics[width=\linewidth]{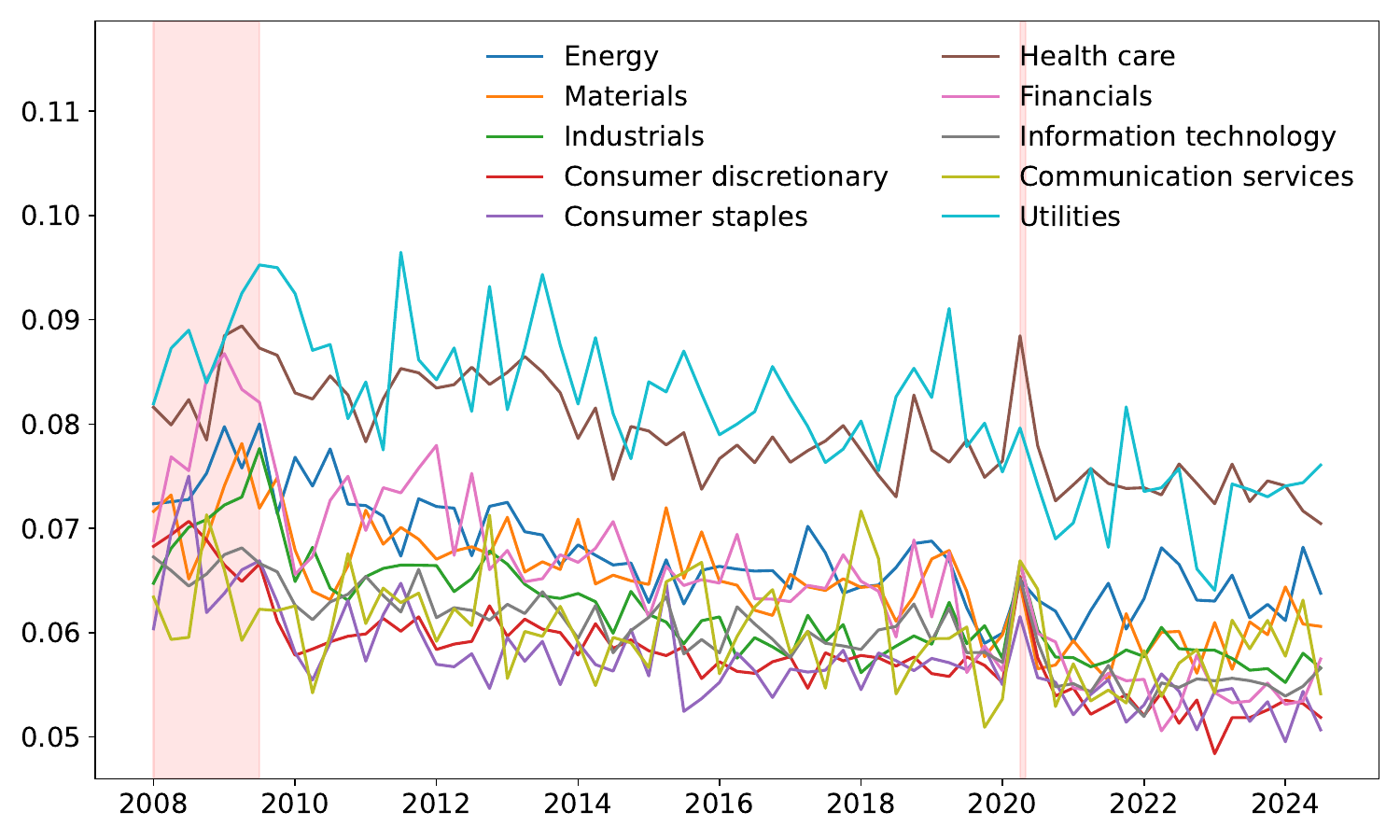}
  \caption{Average Passiveness by Sector \label{fig:timeseries-avg-passiveness-sector}}
  \end{subfigure}
  \caption{Timeseries of Passiveness Scores \label{fig:timeseries}}
  \Description{Timeseries of Passiveness Scores}
\end{figure*}

We begin by examining the time-series behavior of linguistic passiveness. Figure~\ref{fig:timeseries-avg-passiveness} plots the quarterly cross-sectional average passiveness from 2008 to 2024, with the gray shaded region representing the 95\% confidence interval. NBER recession periods are highlighted in red, and the yellow line shows cumulative S\&P 500 returns from January 2005 onward. Two patterns emerge. First, passiveness rises during recessions and periods of market stress: peaks in passiveness coincide with sharp market downturns. During recessions, firm performance is more likely impacted by external forces (credit tightening, demand collapse, supply disruptions). Managers may not be strategically evasive, but they may simply be describing events that were ``done to'' the firm rather than caused by it. Second, average passiveness exhibits a gradual decline over the sample period, consistent with the sustained positive performance of equity markets during our sample period.


Figure~\ref{fig:timeseries-avg-passiveness-sector} reports average passiveness by sector. Utilities (0.082), Health Care (0.078), and Energy (0.068) exhibit the highest levels, suggesting that industry characteristics or sector-specific uncertainty influence linguistic behavior. \citet{rogers2011disclosure} show that optimistic language exposes managers to greater litigation risk; the elevated passiveness in these sectors may therefore reflect a deliberate strategy to reduce litigation exposure given their heightened regulatory scrutiny. Performance in these sectors hinges on external parties: rate approvals in Utilities, drug authorizations in Health Care, and commodity prices in Energy. Because these drivers lie outside managerial control, passive constructions are both more linguistically accurate and more strategically defensible. Given these differences among the sectors, we primarily control for sector fixed effects in subsequent analyses.

Table~\ref{tab:summary-stats} reports summary statistics for passiveness across earnings surprise quintiles, i.e., how well the firm's EPS outperforms the analyst consensus. Quintiles are formed each quarter using the cross-sectional distribution of earnings surprises; the first row reports full-sample averages, and the remaining rows correspond to surprise quintiles. Consistent with intuition, firms with the lowest earnings surprises (Q1) exhibit the highest average passiveness (0.072). Notably, firms with the most positive surprises (Q5) also display elevated passiveness relative to the overall average, suggesting that factors beyond earnings performance influence linguistic style. Comparing the extreme quintiles, firms with poor earnings outcomes are approximately 0.0044 more passive---a 6.6\% increase---with a 95\% confidence interval of $[0.0037, 0.0051]$, or $[5.5\%, 7.6\%]$.


\begin{table}[!htb]
\caption{Summary Statistics}\label{tab:summary-stats}

\centering

\resizebox*{0.8\linewidth}{!}{\begin{tabular}{lcccc}
\toprule
 & Mean & Median & 1\% & 99\% \\
\midrule
Full sample & 0.065 & 0.061 & 0.009 & 0.156 \\
Surprise Q1 & 0.072 & 0.067 & 0.000 & 0.173 \\
Surprise Q2 & 0.065 & 0.061 & 0.011 & 0.153 \\
Surprise Q3 & 0.061 & 0.058 & 0.015 & 0.139 \\
Surprise Q4 & 0.063 & 0.059 & 0.012 & 0.146 \\
Surprise Q5 & 0.067 & 0.063 & 0.000 & 0.167 \\
\bottomrule
\end{tabular}
}
\end{table}

\subsection{Return Prediction}\label{sec:return-prediction}

\begin{figure}[!htb]
  \centering
  \includegraphics[width=0.9\linewidth]{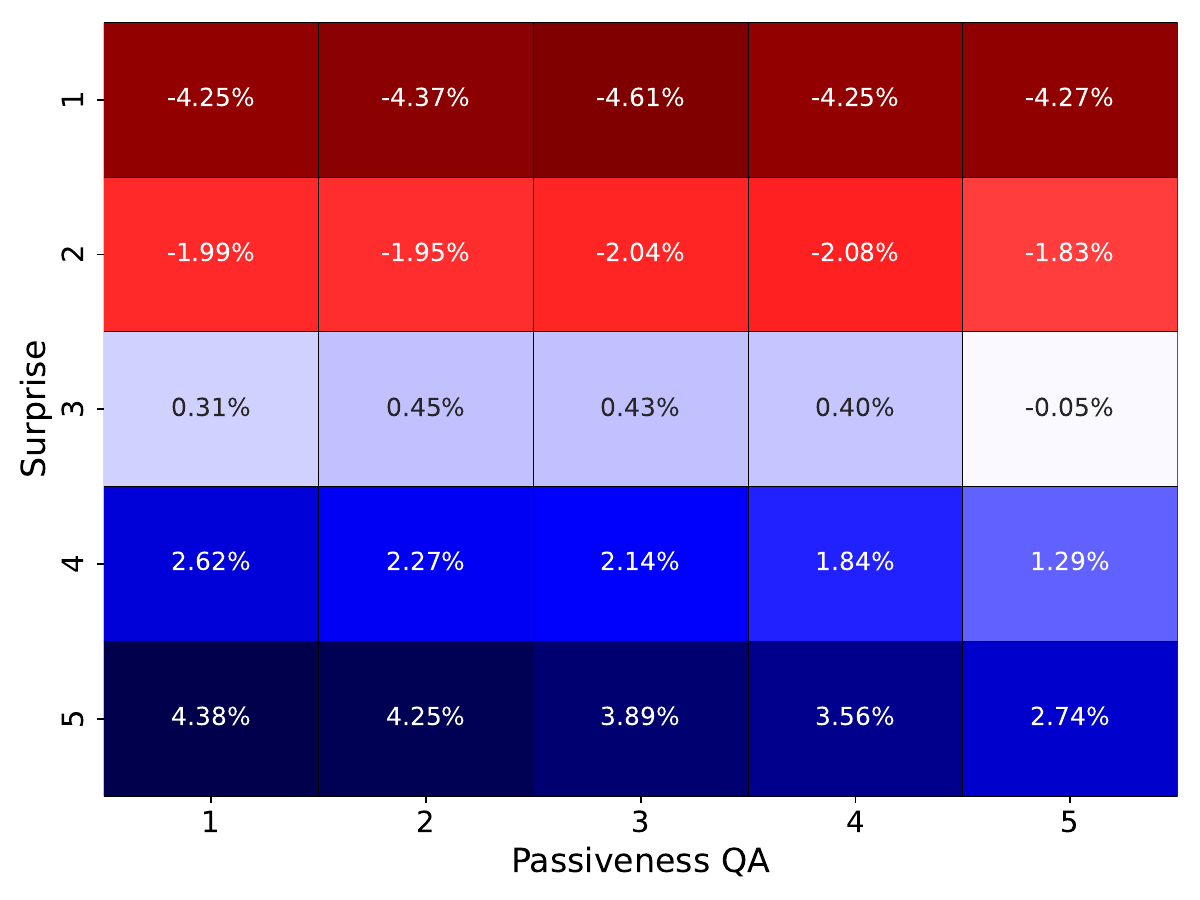}
  \caption{Average Returns Sorted by Passiveness and Surprise}\label{fig:double-sort}
  \Description{Double Sort of Returns on Passiveness and Surprise}
\end{figure}

We next investigate whether passiveness predicts announcement-day returns and whether its effects vary with earnings surprise.
Figure~\ref{fig:double-sort} plots average announcement-day returns sorted into quarterly quintiles of surprise and passiveness.
For firms with low earnings surprises (Q1 and Q2), passiveness has little bearing on announcement-day returns.
In contrast, among firms with high earnings surprises (Q4 and Q5), returns decline substantially as passiveness increases, with the effect strongest in the highest surprise quintile: average announcement-day returns fall from 4.38\% to 2.74\% moving from the lowest to the highest passiveness quintile. 
In untabulated results, we exclude NBER recession periods when computing mean returns per quintile. In that case, mean returns range from 4.23\% to 2.60\% from the lowest to the highest passiveness quintile for the top surprise quintile (Q5) and from 2.57\% to 1.24\% for surprise quintile Q4. 


These numbers suggest that passive language moderates investors' reactions to favorable earnings news. Even when firms outperform analyst expectations, highly passive communication may signal managerial uncertainty, reduced confidence in future performance, or an inclination to attribute positive outcomes to external factors rather than firm-specific actions. Consistent with the attribution literature, investors appear to discount the informativeness of positive earnings surprises when managers simultaneously use passive constructions that distance themselves from the reported results.

To confirm these patterns in a regression framework, Table~\ref{tab:abnormal-return} reports estimates of  the following two regression specifications:
\begin{align}
  \text{AR}_{c,\tau} &= \beta_1 \text{Passiveness}_{c,\tau} + \beta_2 \text{Surprise}_{c,\tau} + \beta_3 \text{Sentiment}_{c,\tau} \nonumber \\ 
  &+ \alpha_\tau + \alpha_s + \epsilon_{c,\tau} \label{eq:ar-1},\\
  \text{AR}_{c,\tau} &= \beta_1 \text{Passiveness}_{c,\tau} + \beta_2 \mathbb{1}_{Surp>0} + \beta_3  \mathbb{1}_{Sent>0} \nonumber\\
  & + \beta_4 \mathbb{1}_{Surp>0}\times \text{Passiveness}_{c,\tau} + \beta_5  \mathbb{1}_{Sent>0}\times \text{Passiveness}_{c,\tau} \nonumber \\ 
  &+ \alpha_\tau + \alpha_s+ \epsilon_{c,\tau} \label{eq:ar-2},
\end{align}
where $\alpha_s$ and $\alpha_\tau$ denote the GICS sector of firm $c$ and the year-quarter fixed effects, respectively. $\mathbb{1}_{Surp>0}$ is an indicator variable that is one if Surprise is positive and zero otherwise, and $\mathbb{1}_{Sent>0}$ is defined analogously for Sentiment. Standard errors are clustered at the quarter and sector levels.

\begin{table}[!htb]
\caption{Announcement-day Abnormal Returns}\label{tab:abnormal-return}

\centering 

\resizebox*{0.95\linewidth}{!}{{
\def\sym#1{\ifmmode^{#1}\else\(^{#1}\)\fi}
\begin{tabular}{@{\extracolsep{2pt}}l*{5}{c}@{}}
\hline\hline

 & (1) & (2) & (3) & (4) & (5) \\
\hline
Passiveness & -0.127\sym{***} & -0.107\sym{***} & 0.016 & 0.039 & 0.049 \\
 & (0.018) & (0.016) & (0.013) & (0.069) & (0.066) \\
Surprise &  & 0.712\sym{***} & 0.680\sym{***} &  &  \\
 &  & (0.074) & (0.065) &  &  \\
Sentiment &  &  & 0.095\sym{***} &  &  \\
 &  &  & (0.011) &  &  \\
$\mathbb{1}_{Surp>0}$ &  &  &  & 0.057\sym{***} & 0.057\sym{***} \\
 &  &  &  & (0.003) & (0.003) \\
$\mathbb{1}_{Sent>0}$ &  &  &  & 0.035\sym{***} & 0.039\sym{***} \\
 &  &  &  & (0.006) & (0.006) \\
$\mathbb{1}_{Surp>0}\times$Passiveness &  &  &  & -0.078\sym{**} & -0.077\sym{**} \\
 &  &  &  & (0.034) & (0.034) \\
$\mathbb{1}_{Sent>0}\times$Passiveness &  &  &  & -0.054 & -0.059\sym{*} \\
 &  &  &  & (0.035) & (0.036) \\
Gunning Fog &  &  &  &  & 0.002\sym{***} \\
 &  &  &  &  & (0.001) \\
Avg words per sentence &  &  &  &  & -0.001\sym{***} \\
 &  &  &  &  & (0.000) \\
Length &  &  &  &  & -0.000\sym{***} \\
 &  &  &  &  & (0.000) \\

\hline
$N$ & 105,642 & 105,642 & 105,642 & 105,642 & 105,642 \\
$R^2$ (\%) & 0.15 & 3.32 & 6.42 & 7.75 & 7.89 \\
\hline\hline
\end{tabular}
}}
\end{table}


Columns (1)--(3) and Columns (4)--(5) report the estimated loadings for the regression \eqref{eq:ar-1} and \eqref{eq:ar-2}, respectively. Columns (1) and (2) report a negative and statistically significant loading on passiveness. When controlling for earnings surprise and sentiment in Column (3), passiveness is no longer statistically significant, indicating it does not explain unconditional cross-sectional variation in announcement-day returns.
However, Columns (4) and (5) indicate that passiveness predicts return variation conditional on positive earnings surprises, in line with the heatmap.
Specifically, Column (4) shows that when interacting passiveness with $\mathbb{1}_{Surp>0}$, there is a negative loading of -0.078 (statistically significant at the 5\% level), indicating that more passive language in conference calls is associated with lower returns among firms with positive earnings surprise. A one standard deviation increase in passiveness is associated with a 0.18\% lower return for those stocks.\footnote{This is the total response to Passiveness and $\mathbb{1}_{Surp>0}\times$Passiveness.} In contrast, when interacting passiveness with $\mathbb{1}_{Sent>0}$, the loading is not statistically significant. We control for additional linguistic characteristics in Column (5), including the Gunning Fog Index \cite{gunning1952}, average words per sentence, and transcript length (word counts). The coefficient on $\mathbb{1}_{Surp>0}\times$Passiveness remains statistically significant with a similar magnitude, indicating that the effect is not driven by readability or document length, confirming robustness. In addition, $\mathbb{1}_{Sent>0}\times$Passiveness becomes statistically significant at the 10\% level.



\subsection{Firm- and Time-Specific Components}

An alternative interpretation of the return predictability in Section~\ref{sec:return-prediction} is that passiveness merely reflects persistent communication styles rather than newly revealed information: the same managers participate repeatedly in conference calls, and speaking patterns may remain stable over time. If so, passiveness would proxy for a fixed managerial trait rather than carry quarter-specific information.
Table~\ref{tab:unexpected-abnormal-return} reports estimates of
\begin{align}
    \text{AR}_{c,\tau} &= \beta_1 \text{Passiveness}_{c,\tau} + \Gamma_t \text{Controls}_{c,\tau} + \epsilon_{c,\tau}.
\end{align}
Standard errors are clustered at the quarter and sector levels. Column (1) reports the univariate relation between passiveness and announcement-day abnormal returns. Columns (2)--(4) progressively introduce firm, sector, and time fixed effects. Specifically, Column (2) includes firm fixed effects based on GVKEY, Column (3) additionally controls for sector fixed effects based on GICS codes, and Column (4) further incorporates time fixed effects.

Across all specifications, the coefficient on passiveness remains negative and highly statistically significant. The estimated coefficient ranges from -0.110 to -0.126, indicating that the relation between passiveness and lower announcement-day abnormal returns is robust to controlling for persistent firm characteristics, industry composition, and time-varying market conditions. The stability of the coefficient across specifications suggests that the predictive power of passiveness is not driven by cross-sectional differences in firms or sectors, nor by common temporal shocks.
This shows that passiveness conveys quarter-specific information about firm performance rather than a static communication trait.

\begin{table}[!htb]
\caption{Announcement Day Abnormal Returns with Effects Control}\label{tab:unexpected-abnormal-return}
\centering 
\resizebox*{0.9\linewidth}{!}{{
\def\sym#1{\ifmmode^{#1}\else\(^{#1}\)\fi}
\begin{tabular}{@{\extracolsep{2pt}}l*{4}{c}@{}}
\hline\hline

 & (1) & (2) & (3) & (4) \\
\hline
Passiveness & -0.117\sym{***} & -0.110\sym{***} & -0.111\sym{***} & -0.126\sym{***} \\
 & (0.017) & (0.019) & (0.019) & (0.020) \\

\hline
$N$ & 105,642 & 105,642 & 105,642 & 105,642 \\
$R^2$ (\%) & 0.14 & 6.88 & 6.85 & 6.89 \\
FirmEffects & N & Y & Y & Y \\
SectorEffects & N & N & Y & Y \\
TimeEffects & N & N & N & Y \\
\hline\hline
\end{tabular}
}}
\end{table}
\subsection{Long-Run Return Predictability}
The announcement-day evidence in Section~\ref{sec:return-prediction} shows that passiveness only has an effect among firms with positive earnings surprises and no significant effect on the announcement day for firms with negative surprises. Does the effect of passiveness on stock returns persist post-earnings announcement? To answer this, we examine whether passiveness predicts post-announcement return drift. Table~\ref{tab:drift} re-estimates the regressions \eqref{eq:ar-1} and \eqref{eq:ar-2}, replacing the response variable with BHAR$[2,61]$, corresponding to cumulative abnormal returns from day $\tau+2$ to day $\tau+61$. Similar to Table~\ref{tab:abnormal-return}, year--quarter and sector fixed effects are included, and standard errors are clustered by sector and quarter.

Columns (1)--(2) report a negative and statistically significant loading (10\% level) on passiveness, but as in Table~\ref{tab:abnormal-return}, when controlling for sentiment, the loading on passiveness becomes statistically insignificant. 
When including Passiveness interacted with the indicator terms $\mathbb{1}_{Surp>0}\times$Passiveness and $\mathbb{1}_{Sent>0}\times$Passiveness, Columns (4)--(5) report that the loading on passiveness is negative and significant at the 5\% and 10\% level, respectively. This indicates that for negative earnings surprise firms, a one standard deviation increase in passiveness now predicts approximately a -0.41\% stock return post-earnings announcement. 
This contrasts with the impact of passiveness on announcement-day stock returns, where the effect of passiveness appears only for positive-surprise firms. Such contrasts can be attributed to investors having a harder time pricing the effect of passiveness in management language when news is negative. 
Columns (4) and (5) further show that for positive earnings surprise firms, the initial response of stock returns on announcement day to passiveness reverses. Precisely, the loadings for $\mathbb{1}_{Surp>0}\times$Passiveness, 0.161 and 0.163 in Columns (4) and (5), respectively, indicate that a one standard deviation increase in passiveness predicts a 0.20\% returns over the [2,61] day window post earnings. This magnitude reverses the negative effect of passiveness on stock returns on the earnings announcement date.


This pattern suggests an asymmetry in how investors process the linguistic signal. When earnings news is positive, investors immediately incorporate passiveness as a moderating signal---a manager who speaks passively about strong results is discounted in real time. When earnings news is negative, however, investors appear to focus on the earnings shortfall itself and pay less attention to the additional information in passive language. The subsequent drift indicates that passiveness carries incremental negative information that is incorporated more slowly into prices, suggesting that there is limited investor attention to linguistic cues when firms' performance is below analysts' expectations.

\begin{table}[!htb]
\caption{Long-term Drift Prediction}\label{tab:drift}

\centering 

\resizebox*{0.95\linewidth}{!}{{
\def\sym#1{\ifmmode^{#1}\else\(^{#1}\)\fi}
\begin{tabular}{@{\extracolsep{2pt}}l*{5}{c}@{}}
\hline\hline

 & (1) & (2) & (3) & (4) & (5) \\
\hline
Passiveness & -0.051\sym{*} & -0.048\sym{*} & 0.004 & -0.136\sym{**} & -0.116\sym{*} \\
 & (0.027) & (0.026) & (0.024) & (0.066) & (0.061) \\
Surprise &  & 0.112 & 0.098 &  &  \\
 &  & (0.114) & (0.115) &  &  \\
Sentiment &  &  & 0.040\sym{***} &  &  \\
 &  &  & (0.007) &  &  \\
$\mathbb{1}_{Surp>0}$ &  &  &  & -0.005 & -0.005 \\
 &  &  &  & (0.006) & (0.006) \\
$\mathbb{1}_{Sent>0}$ &  &  &  & 0.007 & 0.007 \\
 &  &  &  & (0.009) & (0.009) \\
$\mathbb{1}_{Surp>0}\times$Passiveness &  &  &  & 0.161\sym{*} & 0.163\sym{*} \\
 &  &  &  & (0.085) & (0.084) \\
$\mathbb{1}_{Sent>0}\times$Passiveness &  &  &  & 0.001 & 0.007 \\
 &  &  &  & (0.086) & (0.085) \\
Gunning Fog &  &  &  &  & -0.000 \\
 &  &  &  &  & (0.001) \\
Avg words per sentence &  &  &  &  & -0.001 \\
 &  &  &  &  & (0.001) \\
Length &  &  &  &  & 0.000 \\
 &  &  &  &  & (0.000) \\

\hline
$N$ & 105,642 & 105,642 & 105,642 & 105,642 & 105,642 \\
$R^2$ (\%) & 0.00 & 0.01 & 0.08 & 0.02 & 0.03 \\
\hline\hline
\end{tabular}
}}
\end{table}

\subsection{What Drives Managerial Passiveness?}

The previous sections show that passiveness contains information relevant for asset prices. We next investigate what drives passiveness itself. Figure~\ref{fig:timeseries} suggests that passiveness varies in the time-series with macroeconomic conditions and across sectors. We therefore examine how market-, sector-, and firm-level performance measures over the 60 trading days preceding each earnings announcement influence managerial passiveness during conference calls.

We compute the excess market return (EMR$[-60,-1]$) over the 60 trading days preceding the earnings announcement, net of the risk-free rate. Sector performance (SECR$[-60,-1]$) is computed using value-weighted portfolios constructed from 2-digit GICS codes. Following \citet{fama1997industry}, we form initial value-weighted sector portfolios on the first trading day of July each year and subtract the risk-free rate. Firm-level performance is measured using BHAR$[-60,-1]$, the buy-and-hold abnormal return over the same pre-announcement window.  

Table~\ref{tab:drivers} reports estimates of
\begin{equation}
  \text{Passiveness}_{c,\tau} = \alpha + \beta \text{driver}_{c,\tau} + \sum_{i=1}^{10} \gamma_i \mathbb{1}_{\text{sector } i, c, \tau} + \epsilon_{c,\tau},
\end{equation}
where $\mathbb{1}_{\text{sector } i, c, \tau}$ is the sector indicator dummy variable for each firm, and standard errors are clustered by firm and quarter. Column (1) reports the relation between passiveness and contemporaneous surprise and sentiment. Higher earnings surprises and more positive sentiment are each associated with lower passiveness. Together with sector fixed effects, these variables explain approximately 12\% of the variance, leaving a large component unexplained, indicating that passiveness captures information beyond manager sentiment and analyst expectations.
Columns (2)--(5) examine the role of pre-announcement market, sector, and firm performance. Recent market declines have the largest effect on passiveness: a deteriorating market environment in the weeks leading up to the call is associated with more passive language on the day. Firm-level underperformance over the same window similarly predicts more passive communication. Column (5) shows that sector-level performance is largely absorbed by market- and firm-level performance and is not statistically significant in the joint specification.

Taken together, these results suggest that passiveness is not merely a fixed communication style but responds to the information environment facing managers around earnings. When market conditions and firm performance have been adverse, managers tend to be more passive. This is consistent with the attribution and uncertainty mechanisms discussed in Section~\ref{sec:return-prediction}: passive language allows managers to describe outcomes driven by forces beyond their control without explicitly assigning responsibility.

\begin{table}[!htb]
\caption{Passiveness Drivers}\label{tab:drivers}

\centering 

\resizebox*{0.9\linewidth}{!}{{
\def\sym#1{\ifmmode^{#1}\else\(^{#1}\)\fi}
\begin{tabular}{@{\extracolsep{2pt}}l*{5}{c}@{}}
\hline\hline

 & (1) & (2) & (3) & (4) & (5) \\
\hline
Surprise & -0.028\sym{***} &  &  &  &  \\
 & (0.005) &  &  &  &  \\
Sentiment & -0.036\sym{***} &  &  &  &  \\
 & (0.001) &  &  &  &  \\
EMR[-60,-1] &  & -0.006\sym{***} &  &  & -0.008\sym{***} \\
 &  & (0.002) &  &  & (0.003) \\
SECR[-60,-1] &  &  & -0.003\sym{**} &  & 0.002 \\
 &  &  & (0.002) &  & (0.002) \\
BHAR[-60,-1] &  &  &  & -0.001\sym{***} & -0.001\sym{***} \\
 &  &  &  & (0.000) & (0.000) \\

\hline
$N$ & 105,636 & 105,636 & 105,626 & 105,636 & 105,626 \\
$R^2$ (\%) & 11.77 & 7.04 & 7.03 & 7.03 & 7.06 \\
\hline\hline
\end{tabular}
}}
\end{table}

We also examined whether passiveness is associated with the gender composition of conference call participants. Our sample, however, exhibits a substantial gender imbalance: more than 85\% of identified firm participants are male, and female participants account for less than 10\% of Q\&A participants even by 2024. A small fraction of speakers cannot be assigned a gender because their only available identifier is a title prefix (e.g., Dr.) that does not distinguish gender. Given this imbalance, any estimated gender effect would lack statistical power, and we do not pursue a more detailed analysis.

To further validate these patterns, we examine periods of unusually high passiveness in Utilities, Health Care, and Energy---the three sectors with the highest average passiveness. In the Utilities sector, passiveness peaks in 2011Q3 (average 0.095), following the Tohoku earthquake and tsunami in Japan, which generated widespread disruption and uncertainty across global utility markets. The second-highest period, 2013Q3 (0.094), coincides with significant restructuring and consolidation activity in the power and utilities industry. In Health Care, passiveness peaks in 2020Q2 (0.088), corresponding to the COVID-19 pandemic and the uncertainty surrounding its operational and economic consequences. In Energy, passiveness is elevated throughout 2010 (0.073--0.078), in the wake of the Deepwater Horizon disaster, which disrupted oil production and triggered widespread litigation and regulatory scrutiny.

For each of these periods, we identify the Q\&A sessions with the highest passiveness scores and manually examine the extracted passive constructions in context. This qualitative analysis confirms that passive language is most prevalent when managers discuss events they frame as external or beyond their control. Table~\ref{tab:examples} presents three representative examples.

\begin{table*}[!htb]
\caption{Passiveness Examples}\label{tab:examples}

\centering 

\resizebox*{0.95\linewidth}{!}{\begin{tabular}{llll}
\toprule
Sector & Date & Firm & Content\\
\midrule
Utilities & 2011-08-02 & Entergy Corp & The rule it's new, we just heard from Rod and changed significantly from what we saw and expected in 2014.\\
 & & & So we are in the process of actually evaluating what would \textcolor{red}{have to be done} to each of those plants...\\
Energy & 2010-02-26 & TETRA Technologies, Inc & I think, a couple of comments, first, the fourth quarter margins \textcolor{red}{were clearly influenced} by the items I highlighted, 
\\
 & & & which would be the El Dorado startup, as well as some of the delay of some of the Gulf of Mexico projects we have in the backlog, \\
 & & & as well as the \$1.4 million impairment that we wrote off at year end.\\
Health care & 2020-04-30 & Lantheus Holdings & Right now, those uses \textcolor{red}{are suspended}. Because patients \textcolor{red}{aren't being screened} -- \\
 & & & now they're not having the pre hospital screening done for elective procedures because those elective procedures \textcolor{red}{have been postponed}. \\
 & & & Patients aren't going to their physicians' offices for kind of normal ongoing health assessment on a yearly basis. And so those echocardiography studies \textcolor{red}{aren't being ordered}.\\
\bottomrule
\end{tabular}}
\end{table*}

\section{Conclusion}

This paper introduces linguistic passiveness as a measurable dimension of managerial communication in earnings conference calls and shows that it contains information relevant for equity markets. Using 105,642 U.S. conference call events from 2008--2024, we document that passiveness rises during recessions and periods of market stress, and is persistently elevated in Utilities, Health Care, and Energy---sectors characterized by heightened regulatory scrutiny and external uncertainty. Among positive-surprise firms, passiveness negatively predicts announcement-day abnormal returns, with average returns falling from 4.38\% to 2.74\% across passiveness quintiles; this announcement-day discount completely reverses over the subsequent quarter. For negative-surprise firms, passiveness has no announcement-day effect but predicts negative long-run returns, consistent with investor under-attention to linguistic cues when salient bad news dominates the announcement. The return predictability of passiveness is robust to firm, sector, and time fixed effects, reducing concern that it merely reflects a fixed managerial communication style; passiveness is most strongly associated with recent market declines and firm-level underperformance, consistent with managers shifting toward passive constructions when describing outcomes beyond their control. Finally, passiveness retains predictive power after controlling for the \citet{garcia2023colour} sentiment measure and standard readability proxies, suggesting that passiveness constitutes an additional interpretable dimension of managerial communication. By isolating a specific linguistic feature, our measure provides a transparent way to study how managerial language relates to investor reactions and subsequent returns.



\begin{acks}
This research was carried out at Rotman School of Management, FinHub Financial Innovation Lab, University of Toronto. We gratefully acknowledge financial support from the Social Sciences and Humanities Research Council (SSHRC) of Canada grant number 435-2026-0867. We also thank financial support from Royal Bank of Canada Capital Markets.
\end{acks}

\bibliographystyle{ACM-Reference-Format}
\bibliography{main}

@inproceedings{bert,
  address   = {Minneapolis, Minnesota},
  author    = {Devlin, Jacob  and Chang, Ming-Wei  and Lee, Kenton  and Toutanova, Kristina},
  booktitle = {Proceedings of the 2019 Conference of the North {A}merican Chapter of the Association for Computational Linguistics: Human Language Technologies, Volume 1 (Long and Short Papers)},
  doi       = {10.18653/v1/N19-1423},
  editor    = {Burstein, Jill  and Doran, Christy  and Solorio, Thamar},
  month     = {6},
  pages     = {4171--4186},
  publisher = {Association for Computational Linguistics},
  title     = {{BERT}: Pre-training of Deep Bidirectional Transformers for Language Understanding},
  url       = {https://aclanthology.org/N19-1423/},
  year      = {2019}
}

@article{bradley1978self,
  author    = {Bradley, Gifford W},
  journal   = {Journal of personality and social psychology},
  number    = {1},
  pages     = {56--71},
  publisher = {American Psychological Association},
  title     = {Self-serving biases in the attribution process: A reexamination of the fact or fiction question.},
  volume    = {36},
  year      = {1978}
}

@article{doukas2007acquisitions,
  author    = {Doukas, John A and Petmezas, Dimitris},
  journal   = {European Financial Management},
  number    = {3},
  pages     = {531--577},
  publisher = {Wiley Online Library},
  title     = {Acquisitions, overconfident managers and self-attribution bias},
  volume    = {13},
  year      = {2007},
  doi={10.1111/j.1468-036X.2007.00371.x}
}

@inproceedings{ecc-analyzer,
  address   = {New York, NY, USA},
  author    = {Cao, Yupeng and Chen, Zhi and Pei, Qingyun and Lee, Nathan and Subbalakshmi, K. P. and Ndiaye, Papa Momar},
  booktitle = {Proceedings of the 5th ACM International Conference on AI in Finance},
  doi       = {10.1145/3677052.3698689},
  isbn      = {9798400710810},
  location  = {Brooklyn, NY, USA},
  numpages  = {9},
  pages     = {257–265},
  publisher = {Association for Computing Machinery},
  series    = {ICAIF '24},
  title     = {ECC Analyzer: Extracting Trading Signal from Earnings Conference Calls using Large Language Model for Stock Volatility Prediction},
  url       = {https://doi.org/10.1145/3677052.3698689},
  year      = {2024}
}

@misc{finbert,
  archiveprefix = {arXiv},
  author        = {Dogu Araci},
  eprint        = {1908.10063},
  primaryclass  = {cs.CL},
  title         = {FinBERT: Financial Sentiment Analysis with Pre-trained Language Models},
  url           = {https://arxiv.org/abs/1908.10063},
  year          = {2019}
}

@article{garcia2023colour,
  author    = {Garcia, Diego and Hu, Xiaowen and Rohrer, Maximilian},
  journal   = {Journal of Financial Economics},
  number    = {3},
  pages     = {525--549},
  publisher = {Elsevier},
  title     = {The colour of finance words},
  volume    = {147},
  year      = {2023}
}

@article{huang2014tone,
  author    = {Huang, Xuan and Teoh, Siew Hong and Zhang, Yinglei},
  journal   = {The Accounting Review},
  number    = {3},
  pages     = {1083--1113},
  publisher = {American Accounting Association},
  title     = {Tone management},
  volume    = {89},
  year      = {2014}
}

@article{martineau2021rest,
  author  = {Martineau, Charles},
  journal = {Critical Finance Review},
  pages   = {613-646},
  title   = {Rest in peace post-earnings announcement drift},
  volume  = {11},
  year    = {2022}
}

@article{zuckerman1979attribution,
  author    = {Zuckerman, Miron},
  journal   = {Journal of personality},
  number    = {2},
  pages     = {245--287},
  publisher = {Wiley Online Library},
  title     = {Attribution of success and failure revisited, or: The motivational bias is alive and well in attribution theory},
  volume    = {47},
  year      = {1979},
  doi = {10.1111/j.1467-6494.1979.tb00202.x}
}

@inproceedings{earnings1,
author = {Wu, Yuntao and Akin, Ege Mert and Martineau, Charles and Gr\'{e}goire, Vincent and Veneris, Andreas},
title = {Extracting the Structure of Press Releases for Predicting Earnings Announcement Returns},
year = {2025},
isbn = {9798400722202},
publisher = {Association for Computing Machinery},
address = {New York, NY, USA},
url = {https://doi.org/10.1145/3768292.3770344},
doi = {10.1145/3768292.3770344},
booktitle = {Proceedings of the 6th ACM International Conference on AI in Finance},
pages = {596–604},
numpages = {9},
location = {
},
series = {ICAIF '25}
}

@article{passivepy,
author = {Sepehri, Amir and Mirshafiee, Mitra Sadat and Markowitz, David M.},
title = {PassivePy: A tool to automatically identify passive voice in big text data},
journal = {Journal of Consumer Psychology},
volume = {33},
number = {4},
pages = {714-727},
doi = {https://doi.org/10.1002/jcpy.1377},
url = {https://myscp.onlinelibrary.wiley.com/doi/abs/10.1002/jcpy.1377},
year = {2023}
}

@article{Loughran-McDonald-11,
author = {Loughran, Tim and McDonald, Bill},
title = {When Is a Liability Not a Liability? Textual Analysis, Dictionaries, and 10-Ks},
journal = {The Journal of Finance},
volume = {66},
number = {1},
pages = {35-65},
doi = {https://doi.org/10.1111/j.1540-6261.2010.01625.x},
url = {https://onlinelibrary.wiley.com/doi/abs/10.1111/j.1540-6261.2010.01625.x},
year = {2011}
}

@article{mnir,
author = {Matt Taddy},
title = {Multinomial Inverse Regression for Text Analysis},
journal = {Journal of the American Statistical Association},
volume = {108},
number = {503},
pages = {755--770},
year = {2013},
publisher = {Taylor \& Francis},
doi = {10.1080/01621459.2012.734168},
URL = {https://doi.org/10.1080/01621459.2012.734168}
}

@article{passive-1999,
author = {Platow, Michael J. and Brodie, Michelle},
title = {The Effects of Linguistic Voice on Evaluations and Attributions of Ingroup And Outgroup Members},
journal = {Asian Journal of Social Psychology},
volume = {2},
number = {2},
pages = {187-200},
doi = {https://doi.org/10.1111/1467-839X.00033},
url = {https://onlinelibrary.wiley.com/doi/abs/10.1111/1467-839X.00033},
year = {1999}
}

@article{simchon2023agency,
  title={A Computational Text Analysis Investigation of the Relation Between Personal and Linguistic Agency},
  author={Simchon, Almog and Hadar, Britt and Gilead, Michael},
  journal={Communications Psychology},
  year={2023},
  volume={1},
  number={23},
  doi={https://doi.org/10.1038/s44271-023-00020-1}
}

@article{fannes2022crisis,
  title={Shaping Attributions of Crisis Responsibility in the Case of an Accusation: The Role of Active and Passive Voice in Crisis Response Strategies},
  author={Fannes, Gijs and Claeys, An-Sofie},
  journal={Journal of Language and Social Psychology},
  year={2023},
  volume={42},
  number={1},
  pages = {3--30},
  doi={https://doi.org/10.1177/0261927X221108120}
}

@article{fama1997industry,
	title={Industry costs of equity},
	author={Fama, Eugene F and French, Kenneth R},
	journal={Journal of Financial Economics},
	volume={43},
	number={2},
	pages={153--193},
	year={1997},
	publisher={Elsevier}
}

@article{rogers2011disclosure,
	title={Disclosure tone and shareholder litigation},
	author={Rogers, Jonathan L and Van Buskirk, Andrew and Zechman, Sarah LC},
	journal={The Accounting Review},
	volume={86},
	number={6},
	pages={2155--2183},
	year={2011},
	publisher={American Accounting Assocation}
}

@misc{spacy,
author = {Honnibal, Matthew and Montani, Ines and Van Landeghem, Sofie and Boyd, Adriane},
doi = {10.5281/zenodo.1212303},
title = {{spaCy: Industrial-strength Natural Language Processing in Python}},
year = {2020}
}

@book{gunning1952,
  title={The Technique of Clear Writing},
  author={Gunning, R.},
  isbn={9787000014190},
  lccn={51013126},
  url={https://books.google.ca/books?id=ofI0AAAAMAAJ},
  year={1952},
  publisher={McGraw-Hill},
  pages = {36-37},
}


\end{document}